# Charge symmetry breaking in pion-nucleon coupling constant: QCD sum rule approach

J.P. Singh [1] and Alka Upadhyay [2]
Physics Department, Faculty of Science
M.S. University of Baroda, Vadodara-390002, India

ABSTRACT

Using the vacuum-to-pion matrix element of the correlation function of the interpolating fields of the two nucleons, the isospin splitting in the diagonal pion-nucleon coupling constant has been studied in the framework of the conventional QCD sum rule. Some of the implications of this splitting has also been discussed.

[1] E-mail:janardanmsu@yahoo.com
[2] E-mail:alka1977@sify.com

Determination of meson-nucleon couplings is of particular interest in particle physics as well as in nuclear physics. In particle physics, estimate of these parameters is useful to test the low energy behaviour of the QCD. In nuclear physics, nucleon-nucleon interactions are traditionally viewed as arising from meson exchanges. The study of charge symmetry breaking, which is a special case of isospin violation, in pion-nucleon coupling is an important step for investigation of charge symmetry breaking effects in nucleon-nucleon interactions.

Knowledge of these couplings, along with the charge symmetry breaking in them, from QCD may be used for construction of nucleon-nucleon (NN) potential [1]. Introduction of charge symmetry breaking in NN potential models by hand may not be unique. The NN scattering data used in the fitting processes are not precise enough to pick out a specific mesonic channel. Therefore, it is useful to constrain the charge symmetry violation in the pion-nucleon couplings directly from QCD based non-perturbative methods such as QCD sum rule.

At the fundamental level, isospin violation takes place due to charge difference and mass difference of up- and down-quarks. At the phenomenological level, the effect of these differences may get augmented due to strong interaction, and in practice, this may appear in the form of isospin splitting of other phenomenological parameters such as quark condensates. QCD sum rules have been used in past to study isospin breaking in pion-nucleon couplings [2-8]. Three different methods have been used to investigate pion-nucleon coupling constant in the framework of the conventional QCD sum rule. In the three-point function method, one studies the vacuum-to-vacuum matrix element of the correlation function of the interpolating fields of the two nucleons and a meson[3]. However, it has been argued that the method suffers from contamination of higher resonance states.

In the two-point function external field method, one studies the correlation function of the interpolating fields of the two nucleons in the presence of an external pion field [4]. However, the induced condensates appearing in this method are not as reliably known, as the other more commonly used condensates.

In the following we shall follow the third, two-point function method [2,5-7] in which one studies vacuum-to-pion matrix element of the correlation function of the interpolating fields of two nucleons:

$$\prod(p,k) = i \int d^4x \, e^{ipx} \, \langle 0 | T\{\eta(x), \bar{\eta}(0)\} | \pi^0(k) \rangle \qquad \ldots(1)$$

Here, η is the interpolating field of a nucleon and $|\pi^0(k)\rangle$ is the neutral pion state with momentum k. Isospin is suppressed for simplicity. For η, Ioffe's interpolating field will be used. The expression (1) is known to have four Dirac structures.[8]. Among these, the coefficient of the double pole of $i\gamma_5 \hat{p}$ structure on the mass shell vanishes, and the sum rule obtained at the Dirac structure $i\gamma_5$ substantially underestimates the ratio F/D compared to its value known in SU(3) symmetry limit[2].

The sum rule for meson-baryon coupling constant at the structures $i\gamma_5 \hat{p}$ and $\gamma_5\sigma_{\mu\nu}p^\mu k^\nu$ has been studied extensively in [2,5-7]. Kim et al.[5,6] have claimed to find nice features in the sum rule at the $\gamma_5\sigma_{\mu\nu}p^\mu k^\nu$ Dirac structure for calculation of πNN coupling constant. It was observed that for this sum rule the coupling constant comes out to be independent of the choice of the effective Lagrangian, i.e, independent of pseudoscalar and axial vector schemes [7], and is stable against the variation of the continuum parameter due to cancellation of contributions from higher resonances of different parities[5]. We use this sum rule to calculate isospin splitting in the diagonal pion-nucleon coupling constant $g_{\pi NN}$. In the existing result for correlation function (1), we also include quark mass dependent terms. In addition, we also take into account the effect of $\pi^0$–η mixing and electromagnetic correction to the $\pi^0$– quark couplings. In order to reduce the dependence of the splitting in the coupling on the isospin splitting in the quark condensate, which is rather poorly known, we take the ratio of the sum rule for the coupling $g_{\pi NN}$ to the corresponding chiral-odd sum rule for the nucleon mass, and then consider the difference of this ratio for proton and neutron. This resulting sum rule is fitted to a straight line form, which directly gives the difference and the average,

$$\delta g = g^0_{\pi pp} - g^0_{\pi nn}, \qquad g_{\pi NN} = (g^0_{\pi pp} + g^0_{\pi nn})/2 . \qquad \ldots(2)$$

As stated above, in order to construct sum rules for the coupling $g_{\pi NN}$ at the structure $\gamma_5\sigma_{\mu\nu}p^\mu k^\nu$, in addition to the results already derived in Ref.[2], we calculate contributions coming from the quark mass dependent terms of Figs.1(a)-1(b). We enumerate below the Fourier transforms and the Borel transforms of the coefficients of $\gamma_5\sigma_{\mu\nu}p^\mu k^\nu$, of these contributions.

Fig 1(a) $\xrightarrow{F.T.}$ $-(1/2\pi^2)m_d f_\pi \gamma_5\sigma_{\mu\nu} p^\mu k^\nu \ln(-p^2)$ $\xrightarrow{B.T.}$ $(M^2/2\pi^2)m_d f_\pi \gamma_5\sigma_{\mu\nu} p^\mu k^\nu$. $\ldots(3a)$

Fig 1(b) $\xrightarrow{F.T.}$ $-(1/9f_\pi)(m_u/p^4)\langle \bar{u}u \rangle \langle \bar{d}d \rangle \gamma_5\sigma_{\mu\nu} p^\mu k^\nu$

$\xrightarrow{B.T.}$ $(1/9f_\pi)(m_u/M^2)\langle \bar{u}u \rangle \langle \bar{d}d \rangle \gamma_5\sigma_{\mu\nu} p^\mu k^\nu$ $\ldots(3b)$

We have checked that the coefficient of the operator $m_q (\langle \bar{u}u \rangle, \langle \bar{d}d \rangle) \langle (\alpha_s/\pi)G^2 \rangle$ is zero.

So far we have assumed that $\pi^0$ mass eigen state is a pure isovector state. However, it is well known that the mass eigenstates $\pi^0$ and $\eta$ are not pure octet states [9], rather they are mixtures of flavor octet eigenstates $\pi_3$ and $\pi_8$. Denoting $\pi$-$\eta$ mixing angle by $\theta$, the mass eigenstates may be written as:

$$|\pi^0\rangle = |\pi_3\rangle + \theta|\pi_8\rangle, \quad |\eta\rangle = |\pi_8\rangle - \theta|\pi_3\rangle$$

Since $\theta$ is small $\cong 0.01$, this amounts to the replacement for the couplings:

$$g_{\pi^0 pp} = g_{\pi_3 pp} + \theta g_{\pi_8 pp}, \quad g_{\pi^0 nn} = g_{\pi_3 nn} - \theta g_{\pi_8 nn}. \quad \ldots(4)$$

Here, we ignore any possible mixings of $\pi^0$ and $\eta$ with $\eta'$.

It has been pointed out in Ref.[4] that the vertex corrections to $\pi^0 uu$ and $\pi^0 dd$ couplings, due to photon exchanges, can give rise to non negligible charge symmetry breaking in $g_{\pi NN}$. Specifically, it has been found that in the minimum subtraction scheme the following electromagnetic corrections arise to the pion-quark couplings:

$$g_{\pi^0 uu} \to g_{\pi^0 uu}\{1+(\alpha/4\pi)((52/9)-(4/3)\gamma_E)\}, \quad g_{\pi^0 dd} \to g_{\pi^0 dd}\{1+(\alpha/4\pi)((13/9)-(1/3)\gamma_E)\} \quad ..(5)$$

Combining the sum rule for the meson-nucleon couplings as obtained in Ref.[2] at the Dirac structure $\gamma_5 \sigma_{\mu\nu} p^\mu k^\nu$ with the above three types of corrections, we get the following sum rules, after Borel transformation, for the diagonal pion-nucleon couplings:

$$g_{\pi^0 pp}\lambda_p^2(1+D_{\pi p} M^2) = -e^{(m_p^2/M^2)} \frac{\langle \bar{d}d \rangle}{f_\pi} [\{\frac{M^4 E_0(S_0/M^2)}{12\pi^2} + \frac{1}{216}\langle (\alpha_s/\pi)G^2 \rangle\}(1-\frac{\theta f_\pi}{\sqrt{3}f_\eta}) \times$$

$$\{1+\frac{\alpha}{4\pi}(\frac{13}{9}-\frac{1}{3}\gamma_E)\} + f_\pi^2(\frac{4M^2}{3}-\frac{m_0^2}{6}+\frac{26\delta^2}{27})(1+\frac{\theta f_\eta}{\sqrt{3}f_\pi})\{1+\frac{\alpha}{4\pi}(\frac{52}{9}-\frac{4}{3}\gamma_E)\}$$

$$-\frac{M^4}{\langle \bar{d}d \rangle} m_d \frac{f_\pi^2}{2\pi^2} + \frac{\langle \bar{u}u \rangle}{9} m_u], \quad \ldots(6a)$$

$$g_{\pi^0 nn}\lambda_n^2(1+D_{\pi n}M^2) = -e^{(m_n^2/M^2)} \frac{\langle \bar{u}u \rangle}{f_\pi} [\{\frac{M^4 E_0(S_0/M^2)}{12\pi^2} + \frac{1}{216}\langle (\alpha_s/\pi)G^2 \rangle\}(1+\frac{\theta f_\pi}{\sqrt{3}f_\eta}) \times$$

$$\{1+\frac{\alpha}{4\pi}(\frac{52}{9}-\frac{4}{3}\gamma_E)\} + f_\pi^2(\frac{4M^2}{3}-\frac{m_0^2}{6}+\frac{26\delta^2}{27})(1-\frac{\theta f_\eta}{\sqrt{3}f_\pi})\{1+\frac{\alpha}{4\pi}(\frac{13}{9}-\frac{1}{3}\gamma_E)\}$$

$$-\frac{M^4}{\langle \bar{u}u \rangle} m_u \frac{f_\pi^2}{2\pi^2} + \frac{\langle \bar{d}d \rangle}{9} m_d ] . \qquad \ldots(6b)$$

It is clear from the sum rules (6a) and (6b) that the charge splitting in the coupling constant, $\delta g$, has a direct dependence on the isospin splitting of the light quark condensate $\langle \bar{q}q \rangle$ and on the same of the coupling of the nucleon interpolating field, $\lambda_N$. Both these splittings are rather poorly known. However, if we divide these sum rules by the chiral-odd mass sum rules of the respective nucleons, then the $\lambda_N$ dependence will get cancelled and the dependence of $\delta g$ on the isospin splitting of the quark condensate will get minimized. We use the sum rule for the calculation of neutron-proton mass difference derived by Yang et al.[10]. Eliminating $\lambda_p$ of Eq.(6a) with Eq.(17) of [10], and $\lambda_n$ of Eq.(6b) with Eq.(21) of [10] alongwith the electromagnetic mass corrections, we get the sum rules for $g_{\pi^0 pp}$ and $g_{\pi^0 nn}$. Finally, on taking the difference and the average of these two sum rules we get sum rules for $\delta g$ and $g_{\pi NN}$, which we express as

$$\delta g (1+ D^a_{\pi N} M^2) = F_a(M^2), \quad g_{\pi NN}^0 (1+D^s_{\pi N} M^2) = F_s(M^2), \qquad \ldots(7)$$

where $D^a_{\pi N}$ and $D^s_{\pi N}$ are constants. We shall study the sum rule for $g_{\pi NN}^0$ also, in parallel with that for $\delta g$, and compare the result with that derived earlier [2] in a similar approach. Thus a straight line fit of $F_{a,s}(M^2)$ will directly give $\delta g$ and $g_{\pi NN}^0$ in the form of intercepts.

Let us define $a_q = -(2\pi)^2 \langle \bar{q}q \rangle$, $b=\langle g_s^2 G^2 \rangle$, and $\gamma = \frac{\langle \bar{d}d \rangle}{\langle \bar{u}u \rangle} - 1$. Consider the two sets of values of the parameters appearing on the r.h.s of Eq.(7). We have displayed our results for $\delta g$ and $g_{\pi NN}^0$ in Table1. The values of quark masses and quark and gluon condensates of Set I have been taken same as in Refs.[2,10]. Larger values of gluon condensate used in Set II is in accord with the values used in Refs.[11,12]. Smaller values of quark condensate is within the range found in Ref.[12] and the larger values of $\delta^2$ had been used in Ref.[13]. The choice of values of parameters used in Set II is such as to obtain larger value of $g_{\pi NN}$ as claimed in Ref.[2]. $\gamma=-0.01$ has been used in Ref.[3] and $\gamma=-0.00657$ has been used in Ref.[10]. The range of Borel mass squared for Set I is $0.8 \text{GeV}^2 < M^2 < 1.1 \text{GeV}^2$, and for Set II is $0.8 \text{GeV}^2 < M^2 < 1.0 \text{GeV}^2$. This choice is so as to ensure that the contribution of excited states remains less than 50%, and that of the operator of the highest dimension considered remains less than 10% of the total. Moreover, this range is within the ones used in Refs.[2,10]. Also, our continuum thresholds are same as ones used in Refs.[2,10]. We observe that the contributions coming from the non vanishing values of each of $\gamma,\theta,\alpha,\Delta m_q$ and $\Delta m_N$ individually add up almost linearly to give final values of $\delta g$ when all of these parameters are non-zero. We have shown the curves for $F_a(M^2)$ along with the straight line fits in the appropriate ranges for the two sets with all the symmetry breaking parameters being non-zero and with $\gamma = -0.01$ in Figs.(2a) and (2b). To display the quality of the straight line fits, we have computed $\chi^2$ as defined by

$$\chi^2_{a,s} = (1/n) \sum_{i=1}^{n} [F_{a,s}(M_i^2) - P_{a,s}M_i^2 - Q_{a,s}]^2, \quad \chi^2_{1a,s} = (1/n) \sum_{i=1}^{n} [F_{a,s}(M_i^2) - P_{a,s}M_i^2 - Q_{a,s}]^2 / F_{a,s}(M_i^2)$$

where $P_{a,s}$ and $Q_{a,s}$ are the parameters of the straight lines fitting the curves in the specified region. For Set I and II, we get

$$\sqrt{\chi^2_a} = 9.05 \times 10^{-5}, 2.48 \times 10^{-4}; \quad \sqrt{\chi^2_{1a}} = 1.93 \times 10^{-3}, 4.75 \times 10^{-3};$$

$$\sqrt{\chi^2_s} = 1.18 \times 10^{-2}, 1.81 \times 10^{-2}; \quad \sqrt{\chi^2_{1s}} = 1.06 \times 10^{-3}, 1.48 \times 10^{-3}.$$

When all the symmetry breaking parameters are non-zero, we get $\delta g/g_{\pi NN} = -0.008$, and $-0.0092$ for Set I and set II respectively. Using QCD sum rules, in which pion field has been treated as the external field, authors of Ref.[4] have found this ratio as –0.008, and in the cloudy bag model [15] it is –0.006. As is evident from the Table 1, bulk of the contribution to $\delta g$ comes from the nucleon mass difference $\delta m_N$. The quark mass difference and π-η mixing contribute to $\delta g$ in opposite direction, as obtained in [16] also; but these are outweighed by contribution coming from $\delta m_N$. The sign of our result for $\delta g$ differs from that of the three-point function method [3], the chiral bag model [17] and quark gluon model [18]. In the presence of all symmetry breaking parameters, we get the maximum value of $g_{\pi NN}$ as 12.178 which is somewhat lower than the value obtained in a similar approach in Ref.[2]: $g_{\pi NN} \sim 13$-$14$.

Finally, we will discuss some of the implications of the charge symmetry breaking in pion-nucleon coupling constant. Obviously it will contribute to long range part of the charge asymmetric nuclear potential $V_{CA} = V_{nn} - V_{pp}$ for the $^1S_0$ state. In order to calculate its effect on the difference of pp and nn scattering lengths, we use the phenomenological Argonne $v_{18}$ potential [19] disregarding the electromagnetic potential part. With this potential, using $g_{\pi^0 nn}$ and $g_{\pi^0 pp}$, obtained from Set II with all the symmetry breaking parameters present, in the OPEP part of $v_{18}$, we find using the standard method [20] that

$$\left| a_{nn} \right| - \left| a_{pp} \right| \approx 1.0 \text{ fm}$$

consistent with the experimental result [21]:

$$\left| a_{nn} \right| - \left| a_{pp} \right| = (1.5 \pm 0.5) \text{ fm}.$$

Earlier, we had observed that the nucleon mass difference gives the dominant contribution to δg. Reversing the problem, one may ask how much of the nucleon mass difference arises due to δg ? Analysis of the effect of pion loops on nucleon mass has been done by several authors in effective theories of meson-nucleon interaction[22]. Hecht et al. have concluded that the πN-loop reduces the nucleon's mass by ~(10-20)%. Assuming that half of this is due to $\pi^0$-loop, we find that δg will give rise to a mass difference $\delta m_n - \delta m_p \approx$ - (1.5 – 3) MeV, which is a shift in opposite direction to the actual mass difference of the nucleons. Obviously in this case, we cannot neglect the effect of other heavier meson exchanges, and what we have got is far from the end of the story.

**ACKNOWLEDGEMENT**:

Authors gratefully acknowledge the financial assistance from Department of Science and Technology, New Delhi, for this work.

Table 1: The values of $\delta g$ and $g_{\pi NN}$ as defined by Eq.(2) for different values of parameters. Dimensional parameters are in GeV units. The common parameters in the two sets are, So=2.07, $S_{0N}$=2.25, $m_u$= 0.0051, $m_d$ = 0.0089, $m_o^2$ = 0.8, $\mu$= 0.5, $m_p$= 0.9383, $m_n$=0.9396, $f_\pi$= 0.093, $f_\eta/f_\pi$=1.1[14], $\Delta m_q$= 0.0 means $m_u$= $m_d$= 0.007, $\delta m_N$ = 0 means $m_p$= $m_n$= 0.93895 (average nucleon mass) along with the coefficient of $\chi$ in the square bracket of [10] being 5/18 and $m_{em}^2$ = 0. The range of Borel mass squared, $M^2$, is 0.8-1.1. and 0.8-1.0 respectively for Set I and Set II. s.b.p stands for symmetry breaking parameter.

| Parameters | Set I | | Set II | |
|---|---|---|---|---|
| | $a_u$=0.546, b=0.474, $\delta^2$=0.2 | | $a_u$=0.45, b=1.0, $\delta^2$=0.22 | |
| | $\delta g$ | $g_{\pi NN}$ | $\delta g$ | $g_{\pi NN}$ |
| $\gamma = -0.01, \alpha = \theta = \Delta m_q = \delta m_N = 0$ | $-0.0604$ | 9.305 | $-0.0640$ | 12.259 |
| $\alpha = 1/137, \gamma = \theta = \Delta m_q = \delta m_N = 0$ | 0.0135 | 9.294 | 0.0169 | 12.251 |
| $\theta = 0.01, \gamma = \alpha = \Delta m_q = \delta m_N = 0$ | 0.0820 | 9.277 | 0.1031 | 12.229 |
| $\Delta m_q \neq 0, \gamma = \alpha = \theta = \delta m_N = 0$ | 0.0441 | 9.277 | 0.0674 | 12.229 |
| $\delta m_N \neq 0, \gamma = \alpha = \theta = \Delta m_q = 0$ | $-0.1532$ | 9.211 | $-0.2331$ | 12.128 |
| All s.b.p are zero | 0.0000 | 9.277 | 0.0000 | 12.229 |
| All s.b.p.'s are non-zero with $\gamma = -0.01$ | $-0.0754$ | 9.255 | $-0.1116$ | 12.178 |
| All s.b.p.'s are non-zero with $\gamma = -0.00657$ | $-0.0546$ | 9.246 | $-0.0896$ | 12.168 |

**Figure Caption** :

Figs. (1a) and (1b) : The additional diagrams considered in this paper. Cross denotes quark mass insertion.

Fig (2) : The plot of the function $F_a$ of Eq. (7), and the straight line fit $D_a$ used to determine $\delta g$ : (a) for Set I, and (b) for Set II.

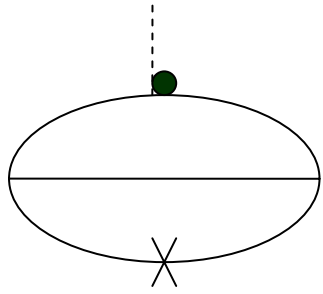

Fig.1(a)

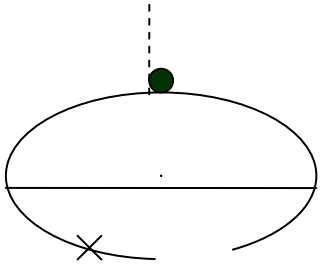

Fig. 1(b)

J.P. Singh.  Phys.Lett.B

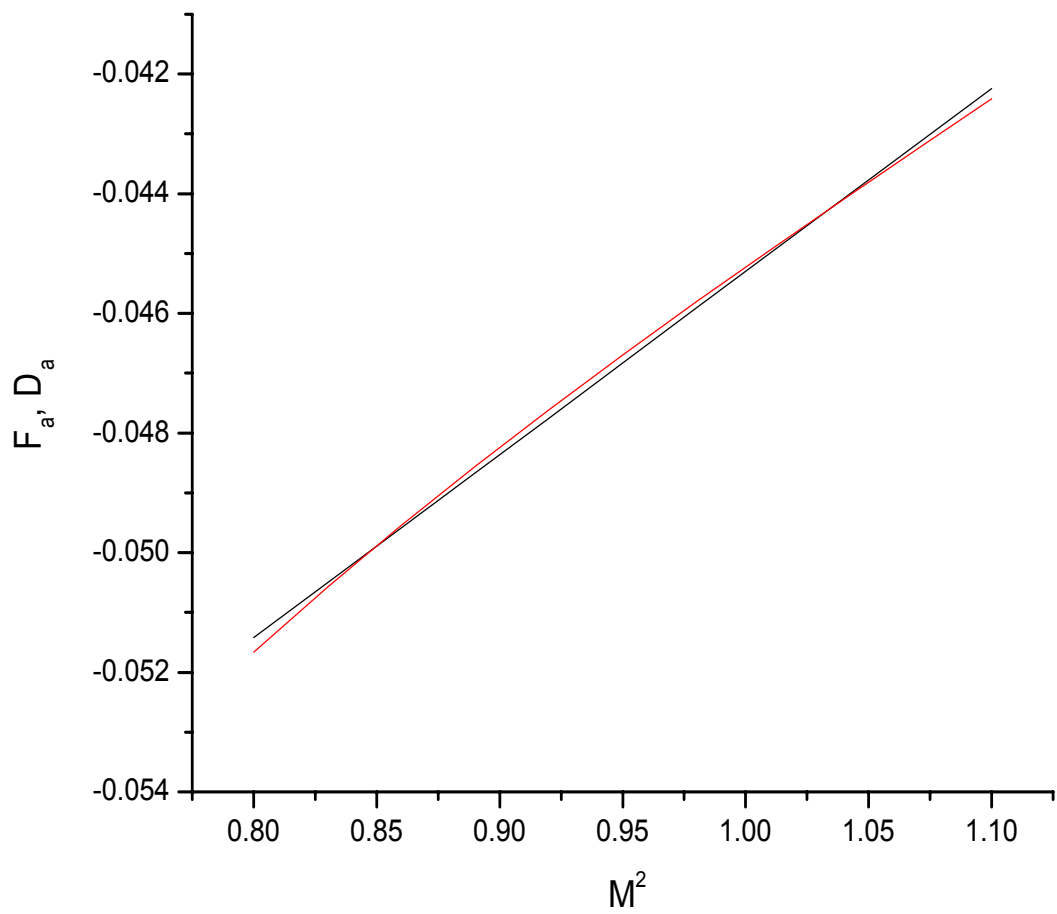

Fig. 2(a)

J.P. Singh. Phys.Lett.B

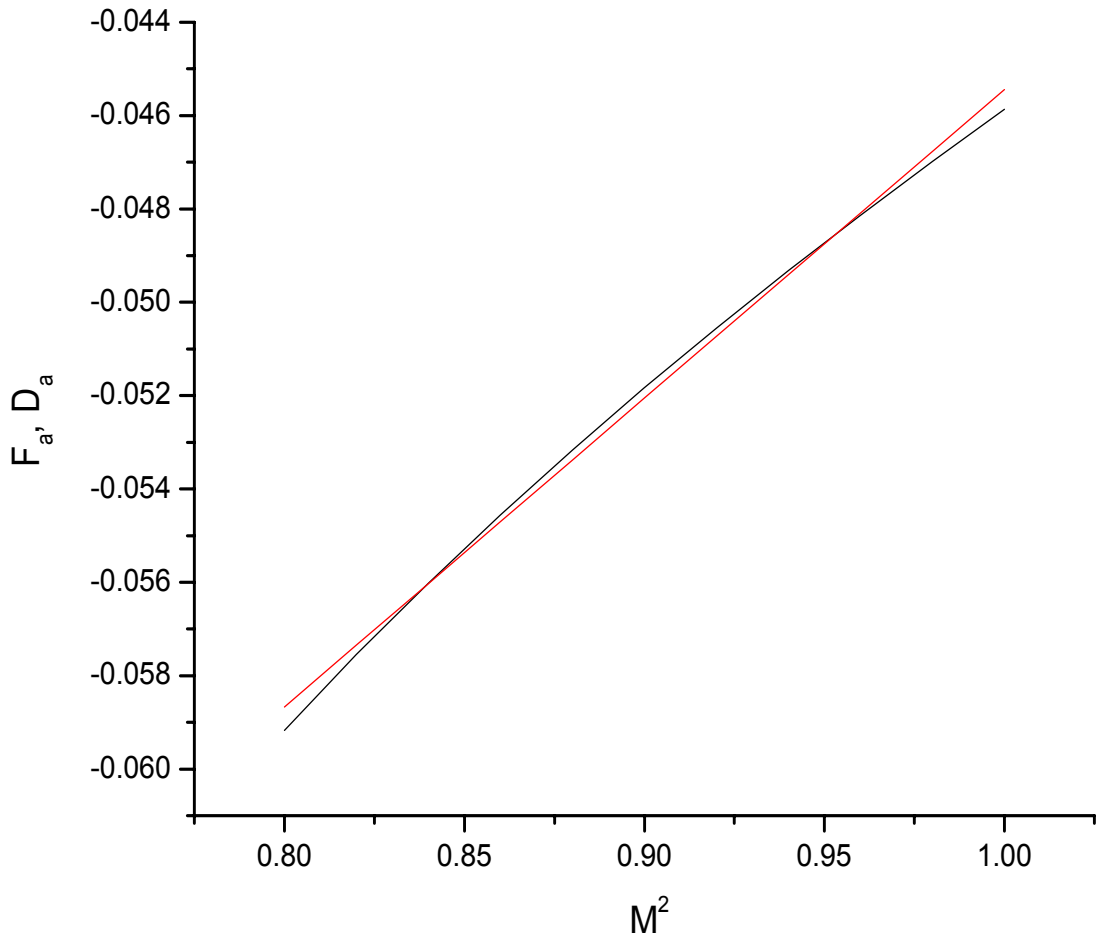

Fig. 2(b)